# Sliding Mode Control Design: a Sum of Squares Approach

Sina Sanjari, *Member, IEEE*, Sadjaad Ozgoli, *Member, IEEE*

*Abstract*—This paper presents an approach to systematically design sliding mode control and manifold to stabilize nonlinear uncertain systems. The objective is also accomplished to enlarge the inner bound of region of attraction for closed-loop dynamics. The method is proposed to design a control that guarantees both asymptotic and finite time stability given helped by (bilinear) sum of squares programming. The approach introduces an iterative algorithm to search over sliding mode manifold and Lyapunov function simultaneity. In the case of local stability it concludes also the subset of estimated region of attraction for reduced order sliding mode dynamics. The sliding mode manifold and the corresponding Lyapunov function are obtained if the iterative SOS optimization program has a solution. Results are demonstrated employing the method for several examples to show potential of the proposed technique.

*Index Terms*—Sliding mode control, Finite time controller, Matched perturbation, Sum of squares (SOS).

## I. INTRODUCTION

Sliding mode control is one of the most effective control methodologies in dealing with a large class of uncertain systems . The controller consists of a high-frequency switching term that completely compensates matched perturbations (i.e. perturbations acting in the direction of control input) [1]. This action takes place when state trajectory remains on the subspace of the state space called "sliding manifold". Much work has been done in the literature to define several sliding mode manifold; the linear sliding manifold is investigated for linear and nonlinear system in [2-4]; nonlinear sliding manifold known as a "terminal sliding mode" also have been introduced in [5] to obtain finite time stability [6]; the problem of singularity of this type of sliding manifold is alleviated and thus "nonsingular terminal sliding mode" have been defined [7]; in order to increase the speed of reaching fast terminal sliding manifold is presented [8].

Several sliding mode manifold have been introduced by many articles [9-11], but selecting a sliding manifold, and determining its parameters is nevertheless an open problem in SMC theory, especially in the case that a complex nonlinear manifold is required. In some applications, linear sliding manifold fails to stabilize the sliding mode dynamics. This difficulty is intensified when finite time stability is an objective of designing control law. While finding parameters of the manifold are possible thorough simulation especially in the case of linear or simple nonlinear manifold, determining such parameters can be tremendously difficult for complex nonlinear manifold that deal with complex nonlinear uncertain systems. On the other hand, it is significant that these parameters are determined so that some objectives are accomplished. Regarding to these objectives and the difficulty of solving, determining sliding manifold leads a subtle problem. The main objective of this article is to design sliding manifold and consequently controller to enlarge the inner bound of the region of attraction in the case of asymptotically stability and finite time stability. The objectives are accomplished systematically using iterative algorithm that search over Lyapunov functions that ensure the type of stability of closed-loop dynamics.

In this paper, we presents a systematic approach utilizing SOS technique, an approach based on Semi-definite programming to deal with polynomial systems [12-14] to obtain sliding mode controller. The approach involves iterative search over sliding manifold and lyapunov function. The proposed method contains SOS optimization program that determine sliding manifold and controller to enlarge the inner bound of region of attraction [15] for the sliding mode dynamics, in case of local stability. Since, pragmatic engineering applications requires finite time stability rather than asymptotic stability, we extend our results to these cases. We introduce a general framework to obtain sliding manifold that ensures finite time stability for sliding mode dynamics. This approach contains all types of terminal sliding mode which are proposed in several papers [5, 16].

The rest of the paper is organized as follows: a brief review On SOS approach and preliminaries are presented in section II. Section III presents the problem and assumptions. Systematic approach to obtain sliding manifold and control is proposed in section IV. In section V, a numerical example is given to demonstrate the effectiveness of the presented method. Finally, section VI concludes the paper.

## II. PRELIMINARIES

This section presents a brief review on SOS decomposition, and other definitions needed to follow the paper.
*Definition 1 (Monomial)*: A monomial $Z_\alpha$ is a function defined as

S. Sanjari is with Tarbiat Modares University, Tehran, IRAN (e-mail: s.sanjari@modares.ac.ir).
S. Ozgoli is with Tarbiat Modares University, Tehran, IRAN. (e-mail: ozgoli@modares.ac.ir).

$$Z_\alpha = x_1^{\alpha_{i_1}} x_2^{\alpha_{i_2}} \dots x_n^{\alpha_{i_n}}$$

For $\{\alpha_{i1}, \dots, \alpha_{in}\} \in Z_+$, and its degree is given by $\deg(Z_\alpha) = \sum_{i=1}^{n} \alpha_i$.

*Definition 2 (Polynomial)*: a real polynomial function $p \in \mathcal{R}[x] = \mathcal{R}[x_1 \dots, x_n]$ is defined as

$$p(x) = \sum_k C_k Z_{\alpha_k}$$

where $C_k \in \mathbb{R}$ and $x \in \mathbb{R}^n$. The polynomial $p(x)$ is said to be of degree $m$ if it corresponds to the largest monomial degree in $p(x)$ i.e. $m = max_k \deg(Z_{\alpha k})$.

In most control problems, "Lyapunov problem" for example, it is important to investigate the non-negativity of polynomials. In general, it is extremely hard or sometimes even impossible to determine non-negativity of such a polynomial. However, a sum of squares polynomials can easily be checked by a SDP. So, in our problem formulation, conditions on non-negativity are replaced by sufficient conditions for polynomials to be SOS.

*Definition 3 (SOS)*: a real polynomial $p(x) \in \mathcal{R}_n$ of degree $d$ is SOS if there exist polynomials such that

$$p(x) = \sum_{i=1}^{r} p_i^2(x)$$

Additionally, the subset of all SOS polynomials in $\mathcal{R}_n$ is denoted by $\Sigma_n$.

The SOS definition implies that the existence of SOS decomposition is sufficient condition for $p(x)$ to be positive semidefinite, i.e. $p(x) \geq 0$. In general, the converse of this result does not hold; however, the possibility of $\mathcal{R}_n$ being $\Sigma_n$ has been calculated in [16]. It is demonstrated that the gap between these two set is negligible.

*Lemma 1 (S-procedure)* [12]: given $\{p_i\}_{i=0}^{m} \in \mathcal{R}_n$, if there exist $\{s_i\}_{i=1}^{m} \in \Sigma_n$ such that $p_0 - \sum_{i=1}^{m} s_i p_i \in \Sigma_n$, then $\bigcap_{i=1}^{m} \{x \in \mathbb{R}^n | p_i(x) \geq 0\} \subseteq \{x \in \mathbb{R}^n | p_0(x) \geq 0\}$.

*Notation:* for matrix $Q \in \mathbb{R}^{n \times n}$, $Q \geq 0$ represents positive semi-definiteness of $Q$; $Q(x) \in \mathcal{R}[x]$ means that $Q(x)$ is a polynomial; $Q(x) \in \mathcal{R}_c[x]$ means that $Q(x)$ is a polynomial contains $c$ variables; $\|a\|$ denotes the 2 norm of $a$.

### III. SYSTEM DESCRIPTION AND PROBLEM STATEMENT

Consider the following nonlinear uncertain system:

$$\dot{x} = f(x) + g(x)u(t) + \xi(t, x) \tag{1}$$

Where $x \in \mathbb{R}^n$ is the state vector, $u \in \mathbb{R}^m$ is the control input, $f(x) \in \mathbb{R}^n$ is a known nonlinear function, and $g(x) \in \mathbb{R}^{n \times m}$ is a known full rank state-dependent matrix. $\xi(x,t) \in \mathbb{R}^{(n-k)}$ is a vector function that models matched perturbation terms. This assumption is not restrictive, and is made by several relevant papers (see [1] for instance). The following assumptions are made on these models.

*Assumption 1*: Although perturbations are considered to be unknown, they are assumed to be bounded i.e.

$$\|\xi(x,t)\| \leq \varphi(x,t) \tag{2}$$

Where $\varphi(x,t)$ is a known function. ∎

*Assumption 2*: System (1) can be represented as a regular form as follows:

$$\begin{cases} \dot{z}_1 = f_1(z) \\ \dot{z}_2 = f_2(z) + L(z)u + \xi_1(t,x) \end{cases} \tag{3}$$

Where $z_1 \in \mathbb{R}^{n-m}, z_2 \in \mathbb{R}^m$, $\xi_1 \in \mathbb{R}^k$, and $\|\xi_1(z,t)\| \leq \varphi_1(x,t)$. ∎

The Sliding manifold is a nonlinear vector function $S(z) \in \mathbb{R}^m$ as

$$S(z) = [M_1(z) \dots M_m(z)]^T \tag{4}$$

Where $S(0, z_2) = 0$ only if $z_2 = 0$. It should be mentioned that there is no restriction for selecting sliding manifold except satisfying the above condition. This condition is necessary since prevent adding new equilibrium to the closed-loop dynamics. In the section V, we define several manifolds that satisfy this condition.

The problem is addressed in this article is to systematically design the sliding manifold and controller so that 1) Stabilize the System (asymptotically, or finite time stable) 2) In the case of local stability that enlarge the inner estimate of region of attraction of closed-loop dynamics.

### IV. MAIN RESULTS

This section investigates a systematic method to design sliding mode manifold and control. Theorem 1, employ sufficient conditions based on SOS constraints that is solvable by SOSTOOLS toolbox [17].

*Theorem 1*: The uncertain system (1) which satisfies assumptions (1-2) will be asymptotically stable by applying the following control law

$$u(t) = -\rho(z,t) \frac{\gamma(S)}{\|\gamma(S)\|} - \left(\frac{\partial S}{\partial z_2} L(z)\right)^{-1} \left(\frac{\partial S}{\partial z_2} f_2(z_1, z_2) + \frac{\partial S}{\partial z_1} f_1(z_1, z_2)\right) \tag{5}$$

$\gamma(s)$ is chosen to be a nonlinear function with $\gamma(S) = 0$ only if $S = 0$. $\rho(z,t)$ is the switching gain function which is selected so that satisfies the following inequality.

$$\rho(z,t) \geq \left(\left\|\frac{\partial S}{\partial z_2}\right\| \varphi_1(z,t) + \eta\right) \tag{6}$$

$\eta > 0$. The sliding manifold $S(z) \in \mathbb{R}^m$ is obtained from the following iterative procedure

1) Set $i=1$, $\beta_1 = -\infty$ and Take small positive constants $\varepsilon_{ij}$ and construct

$$l_k(z) = \sum_{i=1}^{n} \sum_{j=1}^{d} \varepsilon_{ij} z_i^{2j}, \sum_{j=1}^{m} \varepsilon_{ij} > 0, \forall i,j = 1, \dots, n, \varepsilon_{ij} \geq 0, k = 1,2 \tag{7}$$

*Also choose a positive polynomial $p(z)$ to determine the shape of the region of attraction, and select sum of squares polynomial $s_{2_0}$, and a vector polynomial $q_1(z) \in \mathcal{R}^m$ such that $q_1(0) = 0$*

*2) Solve the following SOS program*

---
Find $V \in \mathcal{R}_{n-m}$, $S \in \mathcal{R}_n^m$, $V(0) = S(0) = 0$, $s_1 \in \Sigma_n$ such that

$$V - l_1(z_1) \in \Sigma_{n-m} \tag{8}$$
$$\frac{\partial V}{\partial z_1} f_1(z) - (1-V)s_{2_i} + q_i^T(z)S(z) - l_2 \in \Sigma_n \tag{9}$$
$$-((\beta_i - p)s_1 + (V - 1)) \in \Sigma_n \tag{10}$$

---

*3) Set $V = V_i$, $S = S_i$, and solve the following SOS optimization program*

---
Max $\beta$ over $\beta \in \mathbb{R}$, a vector polynomial $q \in \mathcal{R}_n^m$, and $s_2, s_3 \in \Sigma_n$

$$V_i - l_1 \in \Sigma_{n-m} \tag{11}$$
$$\frac{\partial V_i}{\partial z_1} f_1(z) - (1 - V_i)s_2 + q^T(z)S_i(z) - l_2 \in \Sigma_n \tag{12}$$
$$-((\beta - p)s_3 + (V_i - 1)) \in \Sigma_n \tag{13}$$

---

*4) If $\beta_i > 0$ set $V = V_i$, $S = S_i$. and $\Omega := \{x \in \mathbb{R}^n | V \leq 1\}$ is a subset of region of attraction. if $\beta_i < 0$, set $i = i + 1$, $s_2 = s_{2_i}$, $q^T(z) = q_i^T(z)$ and go to step 2. If $\beta_i = -\infty$, starting from $q_i, s_{2_i}$ does not have any solution.*

*Proof:* In order to prove this theorem, we first show that the control law guarantees sliding mode behavior. Then, by applying the equivalent control method [18], we demonstrate that conditions for asymptotic stability of the sliding mode dynamics are satisfied.

It is straightforward that the controller (5) can maintain the sliding mode, by satisfying reaching condition [1]; therefore, switching gain function satisfying (6) guarantees that the sliding mode can be maintained, $\forall t \in [t_0, \infty)$.

Applying a equivalent control part of the control law (5) on (3) obtain sliding mode dynamic as

$$\begin{cases} \dot{z}_1 = f_1(z) \\ S = 0 \end{cases} \tag{14}$$

Now consider function $V$, the output of the above SOS program as a Lyapunov candidate function. From (8), $V(z) \geq l_1(z)$ where $l_1(z)$ is a positive definite function. $V(0) = 0$ concludes that $V(z)$ is a Lyapunov function. Fixed variable-sized region $p_\beta = \{z \in \mathbb{R}^n | p(z) \leq \beta\}$ is defined in step 1 whose shape is determined by $\beta$. Considering $p(z)$ a positive definite polynomial and maximizing $\beta > 0$ enlarge the inner bound of the set $\Omega$. From (9) and (10) problem can be posed as

$$\text{the set } \{z \in \mathbb{R}^n | V(z) < 1\} \text{ is bounded} \tag{15}$$
$$\{z \in \mathbb{R}^n | p(z) \leq \beta\} \subseteq \{z \in \mathbb{R}^n | V(z) < 1\} \tag{16}$$
$$\{z \in \mathbb{R}^n | V(z) < 1\} \setminus \{0\} \subseteq \{z \in \mathbb{R}^n | \frac{\partial V}{\partial z_1} f_1(z) < 0\} \tag{17}$$

With constraints $S = 0$ where added to the program (14) concludes that for $z_1 = x$ if $x_0 \in \Omega$, then $V(x) \leq V(x_0) \leq 1$. This means that solutions $\varphi^{x_0}$ in $\Omega$ exists and remains inside $\Omega$. Define the set $S_\varepsilon = \{x \in \mathbb{R}^n | \frac{\varepsilon}{2} \leq V(x) \leq 1\}$ for a taken $\varepsilon > 0$. Accordingly $S_\varepsilon \subseteq \Omega \setminus \{0\} \subseteq \{x \in \mathbb{R}^n | \frac{\partial V}{\partial x} f(x) < 0\}$. Since $S_\varepsilon$ is a compact set, $\exists r_\varepsilon > 0$ such that $\dot{V}(x) \leq r_\varepsilon < 0$ on $S_\varepsilon$; therefore, $\exists t^*$ such that $V(x) < \varepsilon$ for all $t > t^*$. This implies that if $x_0 \in \Omega$, then $V(x) \to 0$ as $t \to \infty$.

Let $\varepsilon > 0$, Define $\Omega_\varepsilon = \{x \in \mathbb{R}^n | \|x\| \geq \varepsilon, V(x) \leq 1\}$. $\Omega_\varepsilon$ is compact, with $0 \notin \Omega$. Since $V$ is continuous and positive definite according to (26), $\exists \gamma$ such that $V(x) \geq \gamma > 0$ on $\Omega_\varepsilon$. We have already established that $V(x) \to 0$ as $t \to \infty$. So, $\exists \hat{t}$ such that for all $t > \hat{t}$. $V(x) < \gamma$, and hence, $x \notin \Omega_\varepsilon$, which implies that the origin of system is asymptotically stable for $x_0 \in \Omega$. ∎

The following corollary presents the sliding mode manifold in the case that global asymptotic stability is available, or estimation of the region of attraction is not an objective of designing a controller.

*Corollary 1*: The uncertain system (1) which satisfies assumptions (1-2) will be asymptotically stable by applying the following control law (5), satisfying the condition (6), then The sliding manifold $S(z) \in \mathbb{R}^m$ is obtained from the following iterative procedure

*1) Set $i=1$, $\beta_1 = -\infty$ and Take small positive constants $\varepsilon_{ij}$ and construct (7), and select a vector polynomial $q_1(z) \in \mathcal{R}^m$ such that $q_1(0) = 0$*

*2) Solve the following SOS program*

---
Find $V \in \mathcal{R}_{n-m}$, $S \in \mathcal{R}_n^m$, $V(0) = S(0) = 0$, and $\beta_i > 0$ such that

$$V - l_1(z_1) \in \Sigma_{n-m}$$
$$\frac{\partial V}{\partial z_1} f_1(z) + q_i^T(z)S(z) - \beta_i V \in \Sigma_n$$

---

*3) Set $V = V_i$, $S = S_i$, and solve the following SOS optimization program*

---
Max $\beta$ over $\beta \in \mathbb{R}$, a vector polynomial $q \in \mathcal{R}_n^m$
$$V_i - l_1 \in \Sigma_{n-m}$$
$$\frac{\partial V_i}{\partial z_1} f_1(z) + q^T(z)S_i(z) - \beta V_i \in \Sigma_n$$

---

4) If $\beta_i > 0$ set $V = V_i$, $S = S_i$. if $\beta_i < 0$, set $i = i+1$, $q^T(z) = q_i^T(z)$ and go to step 2. If $\beta_i = -\infty$, starting from $q_i$ does not have any solution.

*Proof:* Using the proof of theorem 1, it is easy to verify that the conclusion hold, and $\dot{V} \leq \beta V$.

The following theorem can be considered as an extension of theorem 1 to guarantee finite time stability instead of asymptotic stability.

*Theorem 2*: The uncertain system (1) which satisfies assumptions (1-2) will be finite time stable by applying the following control law (5). $\gamma(s)$ is chosen to be a nonlinear function with $\gamma(s) = 0$ only if $S = 0$. $\rho(z,t)$ is the switching gain function which is chosen so that satisfies (6). The sliding manifold $S(z) \in \mathbb{R}^m$ is obtained from the following iterative procedure

1) Set $i=1$, $\beta_1 = -\infty$, choose a positive constant $\varepsilon_1$, $c$ and positive integers $r$ and $p$ such that $r > p$, and a polynomial $p(z)$ to determine the shape of the region of attraction, and take a sum of squares polynomial $s_2$, and a vector polynomial $q_1(z)$ where $q_1(0) = 0$. Define slack variable $M = (w(z_1)^T w(z_1))^{\frac{p}{r}}$

2) Solving the following SOS program

---

Find $Q \in \mathbb{R}^{(n-m)\times(n-m)}$, $S \in \mathcal{R}_n^m$, $V(0) = S(0) = 0$, $s_1 \in \Sigma_n$, $K(z) \in \mathcal{R}_n$ such that

$$(Q - \varepsilon_1 I) \in \Sigma \tag{18}$$

$$-(\frac{\partial w}{\partial z_1} f_1(z) Q w(z) + w^T Q f_1(z) \frac{\partial w}{\partial z_1} + cM) + K(z)[M^r - (w^T w)^p] - (1 - w^T Q w)s_{2_i} + q_i^T(z) S(z) \in \Sigma_n \tag{19}$$

$$-((\beta_i - p)s_1 + (w(x)^T Q w(x) - 1)) \in \Sigma_n \tag{20}$$

---

3) Set $Q = Q_i$, $S = S_i$ solving the following SOS optimization program

---

Max $c$ over $c \in \mathbb{R}$, a vector polynomial $q \in \mathcal{R}_n^m$, and $s_2, s_3 \in \Sigma_n$

$$(Q_i - \varepsilon_1 I) \in \Sigma \tag{21}$$

$$-(\frac{\partial w}{\partial z_1} f_1(z) Q_i w + w^T Q_i f_1(z) \frac{\partial w}{\partial z_1} + cM) + K(z)(M^r - (w^T w)^p) - (1 - w^T Q_i w)s_2 + q^T(z) S_i(z) \in \Sigma_n \tag{22}$$

$$-((\beta_i - p)s_3 + (w^T Q_i w - 1)) \in \Sigma_n \tag{23}$$

---

4) If $\beta_i > 0$, set $Q = Q_i$, $S = S_i$, and . and $\Omega := \{x \in \mathbb{R}^n | w(z)^T Q w(z) \leq 1\}$ is a subset of region of reaching if $\beta_i < 0$ set $i = i + 1$, $s_2 = s_{2_i}$, $q^T(z) = q_i^T(z)$ and go to step 2. If $\beta_i = -\infty$ starting from $q_i, s_{2_i}$ does not have any solution.

*Proof:* The proof of this theorem is similar to the theorem 1, so, in the following, we only mention the differences. (18) concludes that matrix $Q$ is positive definite. Defining the Lyapunov function using generalized Krasovskii idea, (18) implies that Lyapunov candidate function is positive definite i.e.

$$V = w(z)^T Q w(z) > 0$$

Taking derivative of the Lyapunov function with respect to time and using (18) and (19) concludes that

$$\dot{V} = f_1^T(z)(\frac{\partial w}{\partial z_1})^T Q w + w^T Q \frac{\partial w}{\partial z_1} f_1(z) \leq c\|w(z_1)\|^{2p/r} \tag{24}$$

since

$$\|w(z_1)\|^2 = (\sqrt{Q}w(z_1))^T Q^{-1}(\sqrt{Q}w(z_1)) \geq \lambda_{min}(Q^{-1})\|(\sqrt{Q}w(z_1))\|^2 = \lambda_{min}(Q^{-1})V \tag{25}$$

It can be easily shown that (24) and (25) yields that $\dot{V} \leq V^\alpha$ for $\alpha = (p/r) < 1$, so the results yields finite time stability. Regarding to proof of theorem 1 the origin of closed-loop dynamics is asymptotically stable for $x_0 \in \Omega$. Now define a function $[0, T(x)) \to (0, V(x)]$, $t \to V(\varphi(x,t))$. Since this function is decreasing and differentiable, its inverse $(0, V(x)] \to [0, T(x))$, $s \to \theta_x(s)$ is differentiable and satisfies the following condition for all $s \in (0, V(x)]$.

$$\theta'_{z_1}(s) = \frac{ds}{-\dot{V}(\varphi(\theta_{z_1}(s), z_1))}$$

Using the variable change $s = V(\varphi(z_1, t))$ concludes that

$$T(z_1) = \int_0^{T(z_1)} dt = \int_{V(z_1)}^0 \theta'_{z_1}(s) ds$$
$$= \int_0^{V(z_1)} \frac{ds}{-\dot{V}(\varphi(\theta_{z_1}(s), z_1))} < +\infty$$

Therefore, settling time is finite and this implies that the origin of the system is finite time stable for $z_{1_0} \in \Omega$. ∎

*Remark 1:* Although selecting a right $w(z)$ maybe in some cases be difficult, but this freedom in selection enables the designer to reflect inside structural information about Lyapunov function in a design approach; therefore, analytic reasoning, and efficient computing could be considered in this design method. For example one choice for $w(z)$ is:

$$w(z) = \begin{bmatrix} 1 & 0 & \cdots & 0 \\ L_{11}(z) & 1 & 0 & 0 \\ \vdots & \cdots & \ddots & 0 \\ L_{n1}(z) & L_{n2}(z) & \cdots & 1 \end{bmatrix} z \tag{26}$$

In order to simplify computing, $w(z)$ can be considered as only $z$, or $f(z)$.

*Remark 2*: While Lyapunov function, which is considered in theorem 2, contains a vast nonlinear functions; nonetheless, it can be extended to general form of nonlinear functions by employing approach have been introduced in [19].

*Remark 3*: Regarding to the nature of the problem of finite time stability, problem statements contains non-polynomial terms; However, SOS approach is presented solely for polynomial vector fields. In order to handle this problem, non-polynomial system must be transmuted into a polynomial one. In recasting procedure, non-polynomial system, which consists of elementary function, is converted to a polynomial system by defining slack variables, although this approach may increase the complexity level of the programming computation [12].

*Remark 4:* Theorems 1 and 2 can be applied also to higher order systems and higher order sliding mode controller by attributing the derivative of appropriate states to new slack variables.

*Corollary 2*: The uncertain system (1) which satisfies assumptions (1-2) will be asymptotically stable by applying the following control law (5), satisfying the condition (6), then the sliding manifold $S(z) \in \mathbb{R}^m$ is obtained from the following iterative procedure

1) Set $i=1$, $c_1 = -\infty$, choose a positive constant $\varepsilon_1$, $c$ and positive integers $r$ and $p$ such that $r > p$, and take a vector polynomial $q_1(z)$ where $q_1(0) = 0$,

2) Solving the following SOS program

---
Find $Q \in \mathcal{R}_{n-m}^m$, $S \in \mathcal{R}_n^m$, $V(0) = S(0) = 0$, and $K(z) \in \mathcal{R}_n$, and $c_i > 0$ such that

$(Q - \varepsilon_1 I) \in \Sigma$
$-(\frac{\partial w}{\partial z_1} f_1(z) Q w + w^T Q f_1(z) \frac{\partial w}{\partial z_1} + c_i M) + K(z)[M^r - (w^T w)^p] + q_i^T(z) S(z) \in \Sigma_n$

---

3) Set $Q = Q_i$, $S = S_i$ solving the following SOS optimization program

---
Max $c$ over $c \in \mathbb{R}$, a vector polynomial $q \in \mathcal{R}_n^m$, and
$(Q_i - \varepsilon_1 I) \in \Sigma$
$-(\frac{\partial w}{\partial z_1} f_1(z) Q_i w + w^T Q_i f_1(z) \frac{\partial w}{\partial z_1} + cM) + K(z)(M^r - (w^T w)^p) + q^T(z) S_i(z) \in \Sigma_n$

---

4) If $c_i > 0$, set $Q = Q_i$, $S = S_i$, if $c_i < 0$ set $i = i + 1$, $q_i^T(z) = q^T(z)$ and go to step 2. If $\beta_i = -\infty$ initializing from $q_i$ does not have any solution.

*Proof:* The proof is straightforward from corollary 1, and theorem 2.

## V. ILLUSTRATIVE EXAMPLES

In this section, one example is provided to show the applicability and flexibility of the method developed in this paper. It should be noted that anywhere needed, the SOS programs are solved by means of SOSTOOLS.

*Example 1*: Consider the following nonlinear system

$$\begin{cases} \dot{x}_1 = x_3 - 2x_1 - x_1^3 - 2x_2^4 x_1 \\ \dot{x}_2 = x_3 - x_2(x_1^2 + x_2^4) \\ \dot{x}_3 = u + \xi(t, x) \end{cases}$$

We use the program in corollary 1 to design SMC. For initialization, we consider $\beta_1 = -100$, $q_1 = x_1 + x_2 + x_3$. One of the sliding manifold that introduce in (4) is

$S = x_3 + B(x_1, x_2)$

Where $B(0,0) = 0$. To simplify the program and the calculation, we consider linear sliding mode manifold. The degree of the Lyapunov function, and $q_i$ are considered two. The results is as follows after 3 iteration

$S = x_3 + 0.66 x_1 + 0.35 x_2$
$V = 0.08 x_1^2 - 0.06 x_1 x_2 + 0.76 x_2^2$
$\beta_4 = 0.32$

For smoothing the control input in simulation, the discontinuous control is approximated by saturation function with $\delta = 0.03$.
In Fig.1 and Fig.2 the simulations of state response, control input, and sliding manifold are presented, which illustrate the stability of system and sliding mode dynamics.

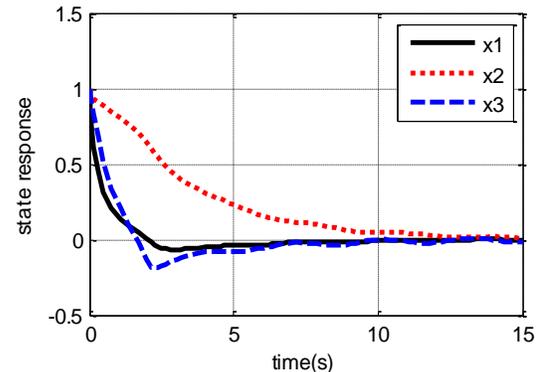

**Fig. 1**. State response of system example 1

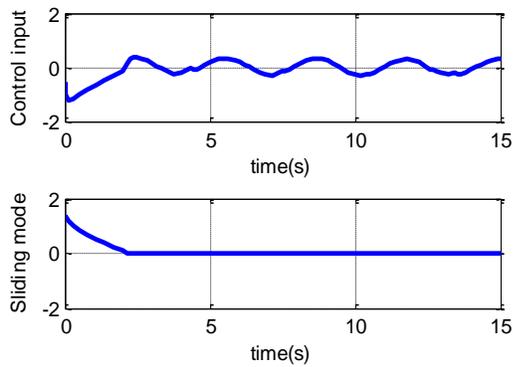

**Fig. 2**. Control input and sliding manifold example 1

## VI. CONCLUSION

A method is presented to design sliding mode control and manifold to enlarge the inner bound of the region of attraction for resulted closed-loop system. Several examples were presented to verify applicability of the proposed method. Benefits of this approach can be summarized: 1) propose a method to design sliding mode control and manifold that results a stable closed loop system 2) provide a systemic approach to establish a sliding mode manifold 3) existence of efficient numerical methods for solving the problem. For further improvement one can investigate the results of theorems to use point-wise maximums of polynomial functions to obtain provable region of attraction of the closed-loop dynamics. In addition, the results can be extended to design sliding mode controller using only output information.